\documentstyle[aps,preprint,eqsecnum]{revtex} \baselineskip 11pt
\begin{document}

\title{ A SUPERSPACE FORMULATION OF AN  ``ASYMPTOTIC" $OSp(3,1|2)$ INVARIANCE
OF YANG-MILLS THEORIES.
 }

%\vspace{.3in}

\author{ Satish D. Joglekar\footnote{e-mail address:- sdj@iitk.ac.in }}
\address{Department of physics \\ Indian Institute of Technology, Kanpur\\
Kanpur, 208016, India \\ and}
\author{Bhabani Prasad Mandal\footnote{ Address for correspondence,    e-mail
:- bpm@tnp.saha.ernet.in}}
\address{
Theory Group,\\ Saha Institute of Nuclear Physics,\\ 1/AF
Bidhannagar, Calcutta- 700 064, India,\\
}

%\vspace{.7in}

\maketitle

%\vspace{.8in}

\begin{abstract}

We formulate a superspace field theory which is shown to be
equivalent to the $ c-\bar{c}$ symmetric BRS/Anti-BRS invariant 
Yang-Mills action. The theory uses a 6-dimensional superspace 
and one $ OSp(3,1|2) $ vector multiplet of unconstrained
superfields. We establish a superspace WT identity and show that
the formulation has an asymptotic $ OSp(3,1|2) $ invariance as the
gauge parameter goes to infinity. We give a physical
interpretation of this asymptotic $ OSp(3,1|2) $ invariance as a
symmetry transformation among the longitudinal/time like degrees
of freedom of $A_\mu $ and the ghost degrees of freedom.

\end{abstract}
%\newpage

\section{INTRODUCTION}

Non-abelian gauge theories are endowed with local gauge invariance
\cite{cheng}. Local gauge invariance leads to relations between
Green's functions of gauge and or ghost fields
collectively denoted by WT identities \cite{iz}. Formulation of
gauge theories in covariant gauges necessities inclusion of
unphysical degrees of freedom corresponding to the longitudinal
and the time like gauge fields. Unitarity of S-Matrix (whenever
defined ) requires that these modes do not contribute to the
intermediate states in the cutting equations \cite{tht}. The
contributions from such intermediate states are canceled by
contributions from diagrams containing ghost intermediate states. This is
demonstrated in gauge theories with the use of the ( on-shell)
WT identities.

Thus the cancellation of intermediate states coming from
longitudinal /time like gauge degrees of freedom and the ghost
degrees of freedom ( we denote this set by R) together is one
of the essential consequence of WT identities. These, in turn, follow
from the BRS symmetry ( or
gauge invariance  )\cite{iz}. This, in turn, suggests that there
should be a formulation of BRS symmetry where the above set of R
of degrees of freedom are explicitly linked together.

There exist many attempts to link $(A, c, \bar{c})$ fields
together. In view of both the commuting and anti-commuting degrees
of freedom involved, this points to a `` Supersymmetric/Superfields
" formulation. A number of superspace /superfield formulation have
been written down which exhibit the BRS symmetry in terms of
translations or rotations in superspace \cite{pr2,sdj}. For a brief
summary of superspace /superfields formulations and their
comparison see comments in Ref. \cite{pro}  and references therein.

The superspace formulation of Ref. \cite{ssp}  constructed superfields $ A(x,
\theta, \bar{ \theta }), c(x, \theta ,\bar{ \theta })$ and $
\bar{c}(x, \theta ,\bar{\theta })$ by hand by ascribing the values
of the additional components $ ( A_{\theta }, A_{\bar{\theta
}}\cdots $ etc) equal to the BRS/Anti-BRS variations \cite{btm} of
these. They exhibited the BRS/Anti-BRS structure thereby. However
as the structure of the superfields was restricted there one could
not construct a full-fledged field theory of these superfields.
The works of references \cite{sdj} and \cite{zpc} (and subsequent works ) attempted
to constructed a field theory of superfields in superspace. Here
the superfields were entirely unconstrained and the superrotations
could be carried out in the formulation. In fact the BRS and Anti-BRS 
were identified with these superrotations and the corresponding WT
identities understood as arising from these. \cite{zpc}. These
constructions had a broken $ OSp(3,1|2) $ symmetry. While these
superspace formulation exhibited the BRS/Anti-BRS structure \cite{sp,zpc}
, and the renormalization  properties \cite{D49} of gauge theories
compactly and correctly. They treated the anti-ghost field asymmetrically
( and as far as we know it is necessary to do this, to exhibit the
renormalization properties in linear gauges.
 Moreover, the underlying $ OSp(3,1|2) $ symmetry was
broken one.

Following the motivations outlined earlier, we  have attempted, in this work, a
formulation that (i)
is a superspace field theories as in Ref. \cite{sdj} (ii) treats gauge, ghost and
anti-ghost fields together in one single supermultiplet. (iii)has an
underlying formal $ OSp(3,1|2) $ symmetry as the basis of construction as a
limiting symmetry of the Lagrange density.
(iv)has WT identities that formally imply that this symmetry becomes exact as gauge
parameter $\eta \rightarrow \infty $ (v) corresponds to the Yang-Mills 
theory in one of its formulation. In fact we find that the superspace
formulation presented here corresponds to the BRS/Anti-BRS invariant
formulation of Baulieu and Thierry-Mieg \cite{btm} with $\beta =1$ ( $ c,
\bar{c} $ symmetric case).

We interpret heuristically the last property in the following manner. We
note ( as done in sec. IIC) that as $ \eta \rightarrow \infty $
the gauge boson propagator is dominated by the longitudinal and time like 
modes. Thus in this limit, the multiplet $(A, c, \bar{c} )$ is dominated by
just the set R of extra modes which enter the unitarity discussion via WT
identities . It is precisely in this limit, the $ OSp(3,1|2) $ symmetry 
is becoming exact.

We now briefly present the plan of the paper. In Sec II, we shall review
the underlying superspace /superfield structure and the OSp group
properties. We briefly discuss the BRS/Anti-BRS symmetric formulation of
Reference \cite{btm} . We also include a brief discussion on the mode structure
of propagator as $\eta\rightarrow \infty $. In section III, we present the
superspace formulation and show its equivalence to the BRS/Anti-BRS
symmetric formulation with $\beta =1$ \cite{btm}. In this section IV, we show
that the generating functional $W[\bar{X}] $ is asymptotically
$(\eta\rightarrow \infty )$ invariant under the $ OSp(3,1|2) $ group. In
sec V, we elaborate on the physical meaning of the result so obtained.

\section{PRELIMINARY}

\subsection{BRS/Anti-BRS symmetric action} 

In this section, we shall review the known results on BRS and anti-BRS 
symmetries of effective action in gauge
theories  \cite{btm}.

We consider the most general effective action in linear gauges given by
Baulieu and Thierry - Mieg \cite{btm} that has
BRS/anti-BRS invariance, when expressed entirely in terms of necessary 
fields $A,c, \bar{c}$ (and no auxiliary fields)
\begin{equation}
S_{eff}[A,c,\bar{c}]=\int \,d^4x \left[-\frac{1}{4}
F_{\mu\nu}^{\alpha}F^{\alpha
\mu\nu} - \sum_{\alpha}\frac{(\partial \cdot A^{\alpha} )^2}{2\eta }-{\cal L}_G \right]
\label{o1}
\end{equation}
with 
%\begin{mathletters}
\begin{eqnarray}
{\cal L}_G & =& (1-\frac{1}{2}\beta)\partial ^\mu \bar{c} D_\mu c + 
\frac{\beta}{2}D^\mu\bar{c}\partial_\mu c
-\frac{1}{2}\beta(1-\frac{1}{2}\beta)
\frac{\eta }{2} g^2[f^{\alpha\beta\gamma}\bar{c}^\beta
 c^\gamma]^2\label{o2}\\
&=&\partial^\mu\bar{c}D_\mu
c+\frac{\beta}{2}gf^{\alpha\beta\gamma}\partial\cdot A^\alpha\bar{c}^
\beta c^\gamma+\frac{1}{8}\beta(1-\frac{1}{2}\beta)\eta 
g^2f^{\alpha\beta\gamma}\bar{c}^\beta\bar{c}^\gamma f^{\alpha\eta\xi}c^\eta c^\xi
\label{o3}
\end{eqnarray}
%\end{mathletters}

 Here we are assuming a Yang- Mills theory with a simple gauge group 
 and introducing the following notations:\\
 \begin{eqnarray*}
\mbox{Lie Algebra:}\;\; [T^\alpha,T^\beta] &=& i f^{\alpha\beta\gamma}
T^\gamma \\
 \mbox{Covariant derivative:}\;\;{(D_\mu c)}^\alpha &\equiv &
D^{\alpha\beta}_
 \mu c^\beta =(-\partial_\mu \delta^{\alpha\beta}+gf^{\alpha\beta\gamma}
A^{\gamma}_\mu)c^\beta\\
 \end{eqnarray*}
 $f^{\alpha\beta\gamma}$ are totally antisymmetric. Note here we have
 changed the  convention for the covariant derivative 
 just to bring it in line
 with notations of Ref.~\cite{sdj}.
This action has the global symmetries
under the following transformations\\
%BRS:\\
%\begin{mathletters}
\begin{eqnarray}
\mbox{BRS :}\nonumber \\
\delta A^{\alpha}_\mu
& =&(D_\mu c)^\alpha \delta \Lambda \nonumber \\
\delta c^\alpha & =&-\frac{1}{2}gf^{\alpha\beta\gamma}c^\beta c^\gamma
\delta\Lambda 
\nonumber\\ 
\delta \bar{c}^\alpha&=& \left(-\frac{\partial\cdot A^\alpha}{\eta 
}-\frac{1}{2}\beta g
f^{\alpha\beta\gamma}\bar{c}^\beta c^\gamma \right)\delta\Lambda\label{o4} \\
\mbox{ and anti-BRS:}\nonumber \\
%\begin{mathletters}
%\begin{eqnarray}
\delta A^{\alpha}_\mu &=& {(D_\mu \bar{c})}^\alpha \delta\Lambda \nonumber \\
\delta \bar{c}^\alpha &=&-\frac{1}{2}gf^{\alpha\beta\gamma}\bar{c}^\beta\bar{c}^
\gamma \delta \Lambda \nonumber\\
\delta c^\alpha &=&\left(-\frac{\partial\cdot A^\alpha}{\eta }
-(1-\frac{1}{2}\beta)g f^{\alpha \beta\gamma}\bar{c}^\beta 
c^\gamma \right)\delta\Lambda\label{o5} 
\end {eqnarray}
%\end{mathletters}
In the anti-BRS transformations the role of $ c$ and $\bar{c} $ are
interchanged in addition to change in some coefficients. Note 
that $\beta=0$ case yields the usual Faddeev- Popov action and
$\beta=1$ yields an action symmetric in $c$ and $\bar{c}$.

\subsection{ Superspace, Superfields and Invariants} 
We shall work in superspace formulation of Yang-Mills theory given in Ref.
[5], which we briefly review in this section. The superspace formulation uses
an underlying six-dimensional superspace described by superspace coordinate
$ \bar{x}^i \equiv (x^\mu, \lambda , \theta )$ with $\lambda , \theta $ are
real Grassmannian variables. Superfields and supersources are function of
superspace coordinates. The superspace is endowed with a metric $g_{ij}$
, with only non zero components
\begin{equation} 
g_{00}= -g_{11}= -g_{22} =-g_{33}= -g_{45}= g_{54} = 1
\end{equation} 
The infinitesimal orthosymplectric coordinate transformations which leaves
the norm of the supervector, $x^ig_{ji} x^j$ invariant are consists of
(i) Six Lorentz transformations which leaves $g_{\mu\nu}x^{\mu}x^{\nu}$
invariant. (ii) Three simplectic transformations which leaves $\lambda
\theta $ invariant and characterized by three infinitesimal parameter.
(iii) and eight SUSY transformations given by
\begin{eqnarray}
x^{\prime \mu} &=& x^\mu + \epsilon ^\mu a \lambda +\delta ^\mu b \theta
\nonumber \\  \lambda ^\prime  &=& \lambda +\delta ^\mu x_\mu b \nonumber
\\ \theta ^\prime  &=& \theta -\epsilon ^\mu x_\mu a
\label{susy}
\end{eqnarray}
[Where $ \epsilon _\mu , \delta _\mu$ are arbitrary four vectors and $a,b$
are real infinitesimal Grassmannians] generated by $S_{4\mu}$ and
$S_{5\mu}$. $S_{4\mu} $ generates transformations with 
 $ \delta =0$  and $ S_{5\mu}$ generates  
transformations with $ \epsilon =0$.
$\frac{ \partial }{\partial x_i}$ are transforms as a covariant vector
under the $ OSp(3,1|2) $ transformations and they are give by
\begin{eqnarray}
\frac{ \partial} { \partial x^{\prime \mu}}
 &=&\frac{ \partial }{\partial 
 x^\mu} +\epsilon
^\mu a \frac{ \partial }{ \partial \theta } -\delta ^\mu b \frac{ \partial
}{\partial \lambda }
\nonumber \\ \frac{ \partial }{  \partial \lambda ^\prime}  &=&\frac{
\partial }{\partial  \lambda} +\epsilon  ^\mu a \frac{ \partial }{\partial 
x_\mu}  \nonumber \\ \frac{ \partial }{\partial \theta^\prime }&=& \frac{ \partial
}{\partial \theta } +  \delta ^\mu b
\frac{ \partial }{\partial x_\mu}
\end{eqnarray}
and the vector superfields $ A_i(\bar{x})\equiv ( A_\mu(\bar{x}) ,c_4
(\bar{x}) , c_5 (\bar{x}) )$ also transform as covariants vectors under
these $ OSp(3,1|2) $ transformations and given by
\begin{eqnarray}
A^{\prime}_\mu &=& A_\mu + \epsilon _\mu a c_5 -\delta _\mu b c_4 
\nonumber \\  c_4 ^\prime  &=& c_4 +\epsilon  ^\mu a A_\mu  \nonumber
\\ c_5 ^\prime  &=& c_5 +\delta ^\mu b A_\mu 
\label{29} 
\end{eqnarray}
The transformations for the vector supersource $ \bar{X}^i (\bar{x}) $ are such
that $\bar{X}^i (\bar{x}) \bar{A}_i (\bar{x}) $ remain invariant under
$ OSp(3,1|2) $ 

We define the scalar product as 
\begin{equation}
A\cdot B = A_i g^{ji}B_j \equiv A^jB_j
\end{equation}
And the tensor invariants are defined as
\begin{equation}
A\cdot B C\cdot D = C_iA_jB_kD_l g^{kj}g^{li} = T_{ij}T_{kl}g^{kj}g^{li}
\end{equation}
where $A^iB_i$ is a commuting quantity.
Using the above definitions of scalar products we construct the following
OSp invariant quantities
$(i)\ F_{ij}F_{kl}g^{kj}g^{li} \ \ (ii)\  \partial ^i[A_i \partial ^jA_j ]
\ \ (iii) \  \partial ^i[A^j \partial _jA_i] \ \ (iv)\  \partial ^i[
 (\partial _i A^j)A_j]$.
Where the superspace field strength tensor, $F_{ij}$ is defined as
\begin{equation} 
F^\alpha _{ij} (\bar{x}) = \partial _i A_j^\alpha (\bar{x}) -A^\alpha _i
\stackrel{\leftarrow}{\partial }_j +gf^{\alpha \beta \gamma }A^\beta _i (\bar{x})
A^\gamma _j (\bar{x}) 
\end{equation} 

\subsection{ Mode structure of gauge propagator}
The propagator in the linear gauges is given by
\begin{equation}
i\Delta_{F\mu\nu} (k,\eta) = \frac{-i}{k^2+i \epsilon }\left [g_{\mu\nu}-
\frac{ k_\mu k_\nu}{k^2+i \epsilon } (1-\eta) \right ]
\end{equation}
We imagine expanding the gauge field (in the momentum space ) in the basis
consisting of the transverse, the longitudinal and the time like degrees of
freedom,
\begin{equation}
A_\mu(k) = \sum_{i=1}^4 \epsilon ^i_\mu(k) a_{(i)} (k^2)
\end{equation}
with  $\epsilon_\mu^{(1)} $ and $\epsilon ^{(2)}_\mu(k) $ are transverse
degrees of freedom with
\begin{equation}
\epsilon _0^{(i)}(k) =0\  , \ \ \ \vec{k}\cdot\vec{\epsilon
}^{(i)}(\vec{k}) =0 \ \ \  i=1,2
\end{equation}
and 
\begin{equation}
\epsilon ^{(3)}_\mu(k) = \left (0, \frac{ -\vec{k}}{|k|} \right );  \ \
\ \epsilon ^{(4)}_\mu = (1,0,0,0)
\end{equation}
We note the orthonormality properties,
\begin{equation}
\epsilon ^{*(i)}\cdot \epsilon ^{(j)} = -\delta ^{ij} + 2 \delta _{i0}
\delta _{j0} \ \ \ \ \  (i,j= 1,2,3,4)
\end{equation}
Then the gauge boson propagator
\begin{equation}
\left < A_\mu(k) \ A_\nu(-k) \right > = \sum_{i,j=1}^4 \epsilon
^{(i)}_\mu (k) \epsilon _\nu ^{(j)}(-k) \left <a_{(i)}(k^2)a_{(j)}(k^2)
\right >
\end{equation}
We recall  
\begin{equation}
\epsilon ^{(i)}_\mu (k) \epsilon_\nu^{( i)}(-k)=-(g_{\mu\nu} -\delta _{\mu 0}
\delta _{\nu 0})+ \frac{ k_\mu k_\nu (1-\delta _{\mu 0})(1-\delta _{\nu
0})}{\vec{k}^2}
\end{equation}
%\end{document}
We then find by comparison,
\begin{eqnarray} 
\left <a_{(i)}(k^2),\ a_{(j)}(k^2) \right > &=& \frac{ \delta _{ij}}{k^2+ i
\epsilon } \ \ \  1\le i,j\le 2 \nonumber \\  
\left < a_j(k^2),\ a_i(k^2) \right >& = &\left <a_{(i)}(k^2),\ a_{(j)}(k^2)
\right > = 0 \ \ \ 1\le i\le 2;\ 3\le j\le 4 \nonumber \\ 
\left <a_{(3)}(k^2),\ a_{(3)}(k^2)\right >&=& -(\eta-1)\frac{|\vec{k}|^2}{
(k^2+i \epsilon )^2}\nonumber \\  
\left <a_{(4)}(k^2),\ a_{(4)}(k^2) \right >&=& -\left [1+\frac{(\eta-1)
|\vec{k}|^2}{k^2+i \epsilon }\right ] \frac{ 1}{k^2+ i \epsilon } \nonumber
\\ 
\left <a_{(3)}(k^2),\ a_{(4)}(k^2) \right > &=& \frac{ (\eta-1)}{2} 
\frac{ |\vec{k}|k_0}{(k^2+i \epsilon )^2}
\end{eqnarray} 
Thus we see that as $ \eta \rightarrow \infty$, the correlation
functions of modes containing $a_{(3)}$ ( the longitudinal ) or
$a_{(4)}$ ( the time like ) go to $\infty, $ while those containing the
transverse components remain unaltered.
We now scale as,
\begin{equation}
a_{(3)} = \sqrt{ \frac{ \eta}{3}}\tilde{a_3};\ \ \ a_{(4)} = \sqrt{ \frac{ 
\eta}{3}}\tilde{a_4};\ \ \tilde{a_i} \equiv a_i \ \ i=1,2
\end{equation}
( The factor of $ \frac{ 1}{3} $ is for future convenience only.)
Then all correlation functions 
$\left < \tilde{a_i}(k^2),\ \tilde{a_j}(k^2) \right >$
have $\eta$- independent limits. Then the expansion of the gauge field reads
\begin{eqnarray} 
A_\mu(k)& =& \sum_{i=1}^2 \epsilon ^{(i)}_\mu(k)\tilde{a}_{(i)}(k^2)+
\sqrt{ \frac{ \eta}{3} } \epsilon ^{(3)}_\mu\tilde{a}_{(3)}(k^2) +
\sqrt{ \frac{ \eta}{3}}\epsilon ^{(4)}_\mu \tilde{a}_{(4)}(k^2) \nonumber
\\ 
&\equiv& A^T_\mu +\sqrt{ \frac{ \eta}{3}}A^L_\mu +\sqrt{ \frac{ \eta}{3}}
A^t_\mu
\end{eqnarray}
The relation above exhibits explicitly the $\sqrt{\eta}$ factors that say
that (after suitable normalization) the longitudinal and the time like
components of a general gauge field become dominant as $\eta
\rightarrow\infty$. This remark will find application in See. V in the 
context of the supermultiplet structure of fields introduced.

\section{ CONSTRUCTION OF SUPERSPACE ACTION}

In this section, we shall present the construction of the superspace action
which is equivalent to the Yang-Mills theory in its BRS/Anti-BRS invariant 
formulation with $\beta =1$ [ See sec. IIA]. The building block of the
superspace action is a covariant vector field $\bar{A}_i(\bar{x}) = \left (
A_\mu (\bar{x}) ,\ c_4 (\bar{x}) ,\ c_5 (\bar{x}) \right ) ( c_4$ and $ c_5
$ will turn out to be related to the antighost field $\bar{c}$ and the
ghost field $c$). We shall also introduce commuting contravariant
vector source $\bar{X}^i (\bar{x}) $ 
. Unlike Ref.  \cite{pro} we don,t however need a scalar superfield and scalar
supersource.
As we shall see later, the Lagrange density turns out to have a graded
structure as the gauge parameter $\eta\rightarrow \infty$
(i) ${\cal L}_0 $ is an OSp invariant action of $O(\eta^0)$
(ii) ${\cal L}_1 $ turns out to be also an OSp invariant, but of
$O(\eta^{-1})$ and (iii) ${\cal L}_2 $ is an Sp(2) invariant symmetry 
breaking term of $O(\eta^{-2})$. Explicitly\footnote{The parameter $\beta$
in ${\cal L}_1$ is not to be confused with $\beta$ in the BRS /anti-BRS
invariant action of Sec. IIA which will always be taken to be 1 in this
work.},
\begin{eqnarray}
{\cal L}_0 &=& \frac{ 1}{4} F_{ij}F_{kl}g^{kj}g^{li} \nonumber \\  
{\cal L}_1 &=& \alpha \partial ^i \left [A_i \partial ^jA_j \right ] +\beta
\partial ^i \left [A^j \partial _j A_i \right ] + \gamma \partial ^i \left
[A^j \partial _i A_j \right ] \nonumber \\  
{\cal L}_2 &=& \frac{\kappa}{2} c^a \partial ^i \partial _i c_a \ \ \ \
a =4,5
\end{eqnarray}
To this we add the source terms
\begin{equation}
{\cal L}_s = \frac{ \partial }{\partial \theta } \frac{ \partial }{\partial 
\lambda }[\bar{X}^i\bar{A}_i ] 
\end{equation}

Under $ OSp(3,1|2) $ transformations ${\cal L}_s$ changes at most by a
total derivative. We then construct the generating functional
\begin{equation}
W[\bar{X}] = \int {\cal D}A \exp{i\int d^4x \left [ {\cal L}_0 +
{\cal L}_1+{\cal L}_2 +{\cal L}_s \right ]}
\end{equation}
where 
\begin{equation} 
{\cal D}\bar{A}\equiv \prod_{i=0}^5{\cal D}A_i (\bar{x}) \equiv \prod_{i=0}^5
{\cal D}A_i(x) {\cal D} A_{i, \lambda }(x) {\cal D}A_{i,\theta }(x)
\end{equation}
In order to establish the equivalence of the above generating functional
with that of the Yang-Mills theory, we carry out the integrations over
the variables $A_{i, \lambda } , A_{i, \theta }$ explicitly as in Ref. 
 \cite{sdj}. The procedure is very straightforward and hence we shall not 
present the details; but only the final result.
Omitting the source terms for the present ( as these are not relevant to
the equivalence ) we find
\begin{equation} 
W[\bar{X}] = \int {\cal D} A_\mu (x) {\cal D} c_4(x) {\cal D}
c_5(x)\exp {i\left [ S_0[A_\mu,c_4,c_5] +\mbox{Source terms} \right ]}
\end{equation}
with\footnote{ $A_{i, \lambda \theta }$ here are not dynamical fields 
\cite{sdj} and can be dropped (i.e. can be set to zero by hand ) in future.}
  omitting redundant terms in $A_{i, \lambda \theta }$.
\begin{eqnarray} 
S_0[A,c_4,c_5] &=& - \frac{ 1}{4} F_{\mu\nu}F^{\mu\nu} - \frac{ (\alpha
-\beta)^2}{2(2 \alpha -\gamma -\beta) }(\partial \cdot A)^2 - \frac{ \alpha
-\beta}{1-2 \gamma } \left [ D_\mu c_4 \partial ^\mu c_5 + \partial _\mu c_4
D^\mu c_5 \right ] \nonumber \\  
&+& \frac{ 2 \gamma }{1-2 \gamma } D_\mu c_4D^\mu c_5 - \frac{ 3(\gamma
+\beta)}{2(\beta+\gamma +1)} (gfc_4c_5)^2-\left [ (\alpha -\beta)^2- \kappa
\right ]\partial _\mu c_4 \partial ^\mu c_5
\label{s0}
\end{eqnarray} 
Comparing with Eq. (\ref{o1}) , we see that $S_0$ of (\ref{s0}) is compatible 
with the action in Eq. (\ref{o1}) only if
\begin{equation}
\mbox{coefficient of } (fA_\mu c_4)(fA_\mu c_5) = 0 \ \ \Rightarrow \gamma =0
\end{equation}
and 
\begin{equation} 
\eta = \frac{ 2 \alpha -\beta}{(\alpha -\beta)^2}
\end{equation}
Further, we use the freedom to define $c_4$ and $c_5$ to set
\begin{equation}
c_5= \frac{ 1}{\sqrt{2(\alpha -\beta)}} c;\ \ \  c_4 = \frac{
1}{\sqrt{2(\alpha -\beta)}} \bar{c}
\label{39}
\end{equation}
then the two actions coincide if, further,
\begin{eqnarray} 
- \frac{ 3\beta}{2(\beta+1)} &=& \frac{ 2 \alpha -\beta}{2} \label{cc} \\ 
  \mbox{  and  }
\kappa & =& (\alpha -\beta)^2 \label{cc1}
\end{eqnarray} 
The quadratic equation of (\ref{cc})  has solutions
\begin{equation} 
\beta = (\alpha +1) \left [ 1\pm \sqrt{1+ \frac{ 2 \alpha }{(\alpha +1)^2}}
\right ]
\label{root}
\end{equation}
Either values of $\beta$ would be acceptable for our purpose.

We shall see in sec. IV that the solution in (\ref{root})  with -ve sign
leads to a superspace Lagrange density that has asymptotic ( i.e. as $\eta\rightarrow
\infty $) symmetry ; and hence we shall make this choice. Thus the
equivalence of the two action with $\beta $ and $ \kappa $ given in terms
of (\ref{cc}) and (\ref{cc1}) is established completely.

We shall, however, be particularly interested in a special case. We further
use the freedom we have in choosing the free parameter $\alpha $ to let 
$0<\alpha \ll 1$ then,
\begin{equation} 
\beta = (\alpha +1) \left [ 1-\sqrt{1+ \frac{ 2 \alpha }{(\alpha +1)^2}
}\right ] \simeq - \frac{ \alpha }{\alpha +1} \simeq - \alpha 
\end{equation} 
Then the gauge parameter becomes 
\begin{equation}
\eta = \frac{ 2 \alpha -\beta}{(\alpha -\beta)^2} \simeq \frac{ 3 \alpha
}{(2 \alpha )^2} = \frac{ 3}{4 \alpha }
\end{equation}
Thus, as $ \alpha \rightarrow 0^+ ,$ our superspace action represents the
BRS/Anti-BRS action with the parameter $\beta$ in \ref{o2} set equal to 1 and $\eta\rightarrow \infty.
$ 
Further,
\begin{equation}
\kappa = ( \alpha -\beta)^2 \simeq 4 \alpha ^2
\end{equation}
expressing all parameters in terms of $\eta (\mbox{ as } \eta \rightarrow \infty
)$, 
\begin{equation} 
-\beta \simeq \alpha \simeq (\frac{3}{4\eta}), \ \ \  \kappa \simeq
4.\frac{9}{16\eta^2} = \frac{9}{4\eta^2}
\end{equation}
and the scaling of (\ref{39}) are re-expressed as
\begin{equation}
c_5 = \sqrt{ \frac{ \eta}{3} } c , \ \ \ \  c_4 = \sqrt{ \frac{ \eta}{3}}
\bar{c}
\end{equation}
To summarize, the superspace action with one free parameter $\eta$
\begin{equation}
\int d^4x \left \{ {\cal L}_0 + \alpha \partial ^i \left [ \partial ^jA_j
A_i- A^j \partial _jA_i \right ] + 4 \alpha ^2 c^a \partial ^i \partial _i
c_a \right \}
\end{equation}
( with $\alpha = \frac{3}{ 4\eta}$ and coefficients valid for $\eta $ large
) is equivalent in the superspace generating functional to the BRS/Anti-BRS 
symmetric action with gauge parameter $\eta (\rightarrow \infty)$ . In the
next section we shall establish the asymptotic OSp invariance for
$W[\bar{K}]$ in other words, the formal equation of the form
\begin{equation} 
\left [W[\bar{X}^\prime  ] - W[\bar{X}]\right ]|_{X^i_{,\theta}=0} = 0( \frac{
1}{\eta})
\end{equation} 

\section{ $OSp(3,1|2)$ WT IDENTITIES}

In this section, we shall consider the consequence of the $ OSp(3,1 2) $
transformations on the source $\bar{X}^i (\bar{x}) $ to obtain the WT identities
for the broken $ OSp(3,1|2) $ symmetry. The result is summarized by the
statement which in effect says that $\eta \rightarrow \infty \ \ W$
recovers $ OSp(3,1|2) $ invariance under the conditions clarified under. It
is also shown how this WT identity embodies exact BRS/anti-BRS symmetry in
the form of the statements \ref{w24a} and \ref{w24b}.

We begin with the generating functional
\begin{equation}
W[\bar{X} (\bar{x}) ] = \int {\cal D} \bar{A}(\bar{x}) \exp \left \{i\int d^4x
\left [{\cal L}_0[\bar{A}]  +{\cal L}_1[ \bar{A}]+{\cal L}_2[\bar{A}]+{\cal L}_s \right ]
\right \}
\label{w}
\end{equation}
We perform an $ OSp(3,1|2) $ 
rotation on the sources $\bar{X}^i$
\begin{equation} 
\bar{X}^i (\bar{x}) \rightarrow \bar{X}^{i\prime}(\bar{x} ) 
\label{w2}
\end{equation} 
with
\begin{equation} 
 \bar{X}^{\prime i} (\bar{x}) = \bar{X}^j(\Lambda^{-1}\bar{x})\Lambda_j^i
\label{w3}
\end{equation} 
Under this transformation, we have the invariance
\begin{eqnarray} 
X^i (\bar{x}) A_i (\bar{x}) &=& X^{\prime i}( \Lambda \bar{x})A^\prime _i
(\Lambda \bar{x}) \nonumber \\  
&=& X^{\prime i} (\Lambda \bar{x})\tilde{\Lambda}_i^jA_j (\bar{x}) 
\label{w4}
\end{eqnarray}
[Here $\tilde{\Lambda}$ is defined in \ref{29}, in particular for
$S_{4\mu}$ and $S_{5\mu}$ transformations ].
Then using  (\ref{w4}), we have  
\begin{eqnarray} 
\int d^4x {\cal L}_s = 
\int \frac{ \partial }{\partial \theta } \frac{
\partial }{\partial \lambda } \left [ \bar{X}^i (\bar{x}) A_i (\bar{x}) \right
]\nonumber \\  
&=& 
\int \frac{ \partial }{\partial \theta } \frac{
\partial }{\partial \lambda } \left [ \bar{X}^{i\prime } (\Lambda\bar{x})
\tilde{\Lambda}^j_i A_j (\bar{x}) \right]
\label{w5}
\end{eqnarray}

In view of the $ SO(3,1) \times Sp(2)$ invariance of the entire $S$, we
expect  new informations to emerge from the transformations associated with
additional supersymmetries $S_{4\mu}$ and $S_{5\mu}.$ Hence we now restrict
ourselves to the $\tilde{\Lambda}$ of Eq. (\ref{29}) given in Sec. IIB. We note
now that 
\begin{equation} 
\Lambda (\bar{x}) = \left ( x^\mu + \epsilon ^\mu a \lambda + \delta ^\mu b
\theta , \lambda + \delta ^\mu b x_\mu , \theta - \epsilon ^\mu a x_\mu
\right )
\label{w6}
\end{equation}
and express
\begin{equation} 
\int d^4x {\cal L}_s =\int d^4x \frac{ \partial }{ \partial \theta } \frac{
\partial }{ \partial \lambda } \left [ \bar{X}^{i \prime }(\bar{x})
\tilde{\Lambda}^j_i A_j (\bar{x}) + \left \{ (\epsilon ^\mu a \lambda +
\delta b \theta ) \partial _\mu \bar{X}^i (\bar{x}) + \delta \cdot x b
\bar{X}^i_{,
\lambda }-\epsilon  \cdot x a \bar{X}^i_{, \theta } \right \} A_i (\bar{x})
\right ]
\label{w7}
\end{equation}
In the last term, we have used the infinitesimal nature of $\epsilon ^\mu
$ and $\delta ^\mu $ to replace $\bar{X}^\prime \rightarrow \bar{X} $ and
$\tilde{\Lambda}\rightarrow 1$. Further,
\begin{eqnarray} 
\int d^4x {\cal L}_s(\bar{X})&-& \int d^4x {\cal L}_s(\bar{X}^\prime )=
\int d^4x \left [ +( \epsilon ^\mu a \lambda + \delta ^\mu b \theta )
\frac{ \partial }{ \partial \theta } \frac{ \partial }{ \partial \lambda  }
(\partial _\mu \bar{X}^i (\bar{x}) A_i (\bar{x}) )\right ]\nonumber \\  &+&
\int d^4x \left [
\epsilon ^\mu a \frac{ \partial }{\partial \theta }(\partial _\mu \bar{X}^iA_i) +
\delta ^\mu b \frac{ \partial }{ \partial \lambda }( \partial _\mu \bar{X}^iA_i)
\right ]+ 
\int d^4x \frac{ \partial }{ \partial \theta } \frac{ \partial }{
\partial \lambda } \left [ \bar{X}^i_{, \lambda \theta } \epsilon \cdot x A_{i,
\theta }-\bar{X}^i_{, \lambda \theta } \delta \cdot x A_{i, \lambda }\right ]
\nonumber \\  &+& \int d^4 x \frac{ \partial }{\partial \theta } \frac{ \partial }{
\partial \lambda } \left [ \bar{X}^\mu( \epsilon _\mu a c_5 - \delta _\mu b c_4)
+ \bar{X}^4(\epsilon ^\mu a A_\mu +\bar{X}^5 \delta ^\mu b A_\mu \right ]
\label{w8}
\end{eqnarray}
Using (\ref{w8}) we can write down change in $W[\bar{X}]$ under an infinitesimal
%\end{document}
OSp transformation (\ref{susy})
\begin{eqnarray} 
\delta W[\bar{X}] && = W[\bar{X}]-W[\bar{X}^\prime ] = << \int d^4x (\epsilon ^\mu a \lambda +
\delta ^\mu b \theta ) \sum_S \partial _\mu S(\bar{x}) 
\frac{ \delta W}{\delta S} \nonumber \\  
&& +i\int d^4x \left [ \bar{X}^\mu _{,\lambda \theta }\left [ \epsilon \cdot x a
A_{\mu, \theta }+ \epsilon _\mu a c_5 - \delta _\mu b c_4 - \delta \cdot x b
A_{\mu, \lambda } \right ] + \bar{X}^4_{ ,\lambda \theta  }\left [\epsilon \cdot x a c_{4,
\theta}+\epsilon ^\mu a A_\mu - \delta \cdot x b c_{4, \lambda } 
\right ] \nonumber\right . \\  
 && \left .  + \bar{X}^5_{, \lambda \theta } \left [ \epsilon \cdot x a c_{5,
\theta }+ \delta ^\mu b A_\mu - \delta \cdot x b c_{5, \lambda } \right ] +
\bar{X}^\mu _{, \lambda } \left [-\delta _\mu b c_{4, \theta }+ \epsilon _\mu a
c_{5, \theta } \right ] - \bar{X}^4_{, \lambda } \epsilon ^\mu a A_{\mu , \theta}
-\bar{X}^5_{, \lambda } \delta ^\mu b A_{\mu, \theta }\right . \nonumber \\  
&& \left . -\bar{X}^\mu_{, \theta } \left [ \epsilon _\mu a c_{5, \lambda
}- \delta _\mu b c_{4, \lambda } \right ] + \bar{X}^4_{, \theta } \epsilon ^\mu a
A_{\mu , \lambda }+\bar{X}^5_{, \theta } \delta ^\mu b A_{\mu , \lambda }+ \left
[ \epsilon^\mu a \partial _\mu \bar{X}^i_{, \theta } + \delta ^\mu b \partial _\mu
\bar{X}^i_{, \lambda } \right ] A_i \nonumber\right . \\  
 && \left . \epsilon^\mu a \partial _\mu \bar{X}^\nu A_{\nu , \theta } +
\delta ^\mu b \partial _\mu \bar{X}^\nu A_{\nu , \lambda } - \epsilon^\mu a
\partial _\mu (\bar{X}^4c_{4, \theta }+\bar{X}^5 c_{5, \theta }) - \delta ^\mu b
\partial _\mu (\bar{X}^4c_{4, \lambda }+\bar{X}^5c_{5, \lambda }) \right ] >>
\label{w9}
\end{eqnarray} 
Here we have dropped terms proportional to $A_{, \lambda \theta }^i$ ( as
these fields can be set to zero). The double bracket, $<< >> $ has been
used to denote that the expression inside it is actually inside the path
integral.

We now evaluate $\delta W[\bar{X}]$ for the``supersymmetry transformations "
$ S_{4\mu}$ only; i.e. set $ \delta =0$. We, further, note that the
sources $-\bar{X}^\mu_{, \theta }$  are to generate the Green,s functions of the
composite operator involved in the anti-BRS  transformations . These are
not required to evaluate the basic Green,s functions of the Yang-Mills
theory, nor the BRS WT identities. Hence, we evaluate (\ref{w9}) at $\bar{X}^\mu
_{, \theta } =0$. [ The spurious terms involving $\partial _\mu \bar{X}^i$ can
also be set to zero as $\bar{X}^i$ are  sources for $A_{, \lambda \theta }$
 which are redundant  field].

Now, the first term on the right hand side of (\ref{w9}) ( here $\sum_S$ goes
over the sources $\bar{X}^i; \bar{X}^i_{, \lambda }; \bar{X}^i_{, \theta
};\bar{X}^i_{,\lambda
\theta}$) vanishes by the translational invariance of $W[\bar{X}]$ in $x^\mu$.

We organize the rest of the terms in $ \delta W$ as
\begin{eqnarray} 
\delta W[\bar{X}]|_{\bar{X}^i_{, \theta  }=0=\bar{X}^i(x)} &=& <<i \int d^4x \left
\{ \bar{X}^\mu_{,
\lambda \theta }\left ( \epsilon\cdot x a A_{\mu, \theta }+ \epsilon _\mu a
c_5 \right ) +\bar{X}^4_{, \lambda \theta } \left ( \epsilon\cdot x a c_{4,
\theta }+ \epsilon^\mu a A_\mu \right ) \right . \nonumber \\  
&& \left . + \bar{X}^5_{ ,\lambda \theta }(\epsilon\cdot x a c_{5, \theta
}) + \bar{X}^
\mu_{, \lambda }\epsilon _\mu a c_{5, \theta } - \bar{X}^4_{, \lambda } \epsilon
^\mu aA_{\mu , \theta } \right \} >>
\label{w10}
\end{eqnarray} 
We shall simplify the expression on the right hand side employing the 6-D
gauge invariance of ${\cal L}_0$ \cite{bpm}. We consider the gauge
transformations 
\begin{equation} 
\delta A_i = D_i( c_5 \epsilon\cdot x a)
\label{w11}
\end{equation} 
and the consequent transformations 
\begin{equation} 
\delta A_{i,\theta } = \frac{ \partial }{ \partial \theta } 
\left [ D_i( c_5 \epsilon\cdot x a) \right ]; \ \ \ 
\delta A_{i,\lambda  } = \frac{ \partial }{ \partial \lambda  } 
\left [ D_i( c_5 \epsilon\cdot x a) \right ]
\label{w11a}
\end{equation}
Under (\ref{w11}) and (\ref{w11a}) ; ${\cal L}_0$ is gauge invariant
\begin{equation} 
\int d^4x \left [ \delta A_i \frac{ \delta S_0}{\delta A_i} + \delta A_{i,
\theta } \frac{ \delta S_0}{ \delta A_{i, \theta} } + \delta A_{i, \lambda }
\frac{ \delta S_0}{\delta A_{i, \lambda }} \right ]=0
\label{w12}
\end{equation} 

We now invoke the equations of motion
\begin{eqnarray} 
\frac{ \delta S_0}{\delta A_\mu}&=& -\bar{X}^\mu_{, \lambda \theta }+ (\alpha
-\beta) \partial _\mu(c_{4, \theta }-c_{5, \lambda }) \nonumber \\  
\frac{ \delta S_0}{\delta c_4}&=& \bar{X}^4_{, \lambda \theta }+ (\alpha
-\beta) \partial ^\mu(A_{\mu, \theta })+\kappa \partial ^2 c_5 \nonumber \\  
\frac{ \delta S_0}{\delta c_5}&=& \bar{X}^5_{, \lambda \theta }- (\alpha
-\beta) \partial ^\mu A_{\mu, \lambda  }-\kappa \partial ^2 c_4\nonumber \\  
\frac{ \delta S_0}{\delta A_{\mu , \lambda } }&=& \bar{X}^\mu_{, \theta }- (\alpha
-\beta) \partial _\mu c_5  \nonumber \\  
\frac{ \delta S_0}{\delta c_{4, \lambda }}&=& \bar{X}^4_{, \theta }- 2
\beta  c_{5, \theta } \nonumber \\  
\frac{ \delta S_0}{\delta c_{5, \lambda }}&=& \bar{X}^5_{, \theta }+ (\alpha
-\beta) \partial\cdot A-( \alpha -\beta) c_{5, \lambda }+ 2 \alpha c_{4,
\theta } \nonumber \\  
\frac{ \delta S_0}{\delta A_{\mu , \theta }}&=& -\bar{X}^\mu_{, \lambda}+ (\alpha
-\beta) \partial _\mu c_4  \nonumber \\  
\frac{ \delta S_0}{\delta c_{4, \theta }}&=&- \bar{X}^4_{, \lambda }+ (\alpha
-\beta) \partial\cdot A-( \alpha -\beta) c_{4, \theta }+ 2 \alpha
c_{5,\lambda } \nonumber \\  
\frac{ \delta S_0}{\delta c_{5 , \theta }}&=& -\bar{X}^5_{, \lambda}+ 
2\beta c_{4, \lambda }    
\label{w13}
\end{eqnarray} 
[It is understood that these Eqs are in double brackets.]

Using (\ref{w13}) in (\ref{w12}), we obtain
\begin{eqnarray} 
<< -i\int d^4x && \left [ D_\mu(\epsilon \cdot x a c_5) \bar{X}^\mu_{, \lambda
\theta } -D_4(\epsilon\cdot x a c_5)\bar{X}^4_{, \lambda \theta }
-D_5(\epsilon\cdot x ac_5)\bar{X}^5_{, \lambda \theta } + \frac{ \partial
}{\partial \theta }[D_\mu(\epsilon\cdot x ac_5)]\bar{X}^\mu_{, \lambda }\nonumber
\right .\\ && \left . \frac{ \partial }{\partial \theta }[D_4 \epsilon
\cdot x a c_5 ]
\bar{X}^4_{, \lambda }+ \frac{ \partial
}{\partial \theta }[D_5(\epsilon\cdot x a c_5)]\bar{X}^5_{, \lambda } \right ]
\nonumber \\   &&  = O( \alpha , \bar{X}^i_{, \theta })
\label{w14}
\end{eqnarray} 

We now subtract (\ref{w14}) from (\ref{w10}) to obtain
\begin{eqnarray} 
\delta W[\bar{X}] |_{\bar{X}^i_{, \theta} =0=\bar{X}^i}= << i\int d^4x \epsilon\cdot x a \left
\{ (A_{\mu, \theta }+D_\mu c_5) \bar{X}^\mu_{, \lambda \theta}-(c_{4, \theta
}-D_4c_5)\bar{X}^4_{, \lambda \theta }-(c_{5, \theta }-D_5c_5)\bar{X}^5_{, \lambda
\theta } \nonumber \right .\\ \left . 
-\bar{X}^4_{, \lambda \theta } \epsilon ^\mu a A_\mu - \frac{ \partial }{
\partial \theta }(D_\mu c_5)\bar{X}^\mu_{, \lambda }+ \frac{ \partial }{\partial
\theta }(D_4c_5) \bar{X}^4_{, \lambda }+ \frac{ \partial }{\partial \theta
}(D_5c_5)\bar{X}^5_{, \lambda } -\bar{X}^4_{, \lambda } \epsilon ^\mu a A_{\mu , \theta
}\right \}>>
\label{w15}
\end{eqnarray} 
Now we recall the equation of motion
\begin{eqnarray}
 << A_{\mu , \theta }+D_\mu c_5 + (\alpha -\beta)\partial _\mu c_5 -
\bar{X}_{,
\theta }^\mu>> =0 \label{w16a}\\
<< c_{5, \theta } - D_5c_5 + 2\beta c_{5, \theta } -\bar{X}^4_{, \theta} >> =0
\label{w16b}
\end{eqnarray} 
and 
\begin{equation} 
\frac{ \partial }{\partial \theta }(D_5c_5) = -2f^{\alpha
\beta\gamma}c^\beta_{5, \theta }c^\gamma_5 =0
\label{w17}
\end{equation} 
Which can be obtained by using (\ref{w16b}) at $\bar{X}^4_{, \theta }=0$. 
We further have the equation of
motion of $c_{4, \theta }$ and $c_{5, \lambda }$.
\begin{eqnarray}
<<c_{4, \theta } + c_{5, \lambda } + g fc_4c_5 +\beta(c_{4, \theta } + c_{5,
\lambda }) - \frac{ \bar{X}^4_{, \lambda }-\bar{X}^5_{, \theta }}{2}>> =0
\label{w16c} \\
<<(2 \alpha -\beta)(c_{4, \theta }-c_{5, \lambda }) + (\alpha -\beta)
\partial \cdot A- \frac{ \bar{X}^5_{, \theta }-\bar{X}^4_{ ,\lambda}}{2}>>=0
\label{w16d}
\end{eqnarray} 
Subtracting (\ref{w16d}) from (\ref{w16c}) and setting $\bar{X}^5_{, \theta }$ we
obtain
\begin{eqnarray} 
<< c_{4, \theta }&-&D_4c_5>>|_{\bar{X}^5_{, \theta }=0}\nonumber \\  
&& = -\beta<<c_{4, \theta
}+c_{5, \lambda }>>+ (2 \alpha -\beta)<<c_{4, \theta} - c_{5, \lambda }>>+
(\alpha -\beta) << \partial \cdot A>> = O( \alpha )
\label{w18}
\end{eqnarray} 
Further, using (\ref{w16a}) and (\ref{w16b}), we obtain ( at $ \bar{X}^i_{, \theta }
=0$ )
\begin{equation} 
\frac{ \partial }{\partial \theta }(D_\mu c_5) = O( \alpha )
\label{w19}
\end{equation}
[ We recognize in (\ref{w17}) and in (\ref{w19})the usual BRS invariance
statement of $\frac{ 1}{2} fcc$ and $ D_\mu c.$ ]

We further recall the equation of motion of $c_4$
\begin{eqnarray}
<<-D_\mu^{\beta \alpha } (A_{\mu , \theta } +D_\mu c_5)^\alpha + f^{\alpha
\beta \gamma }c^\gamma _4(2c_{5, \theta }^\alpha +gf^{\alpha \eta \delta
}c_5^\eta c_5^\delta )-gf^{\alpha \beta \gamma }c_5^\gamma (c_{4, \theta
}^\alpha +c_{5, \lambda }^\alpha +gf^{\alpha \eta \delta }c_4^\eta
c_5^\delta ) \nonumber \\  
-\bar{X}^4_{, \lambda \theta } -( \alpha -\beta) \partial ^\mu A_{\mu , \theta }
+\kappa \partial ^2 c_5 >> =0
\label{w20}
\end{eqnarray}
On account of (\ref{w16a}),(\ref{w16b}) and (\ref{w18}) used successively in the
left hand side of (\ref{w20}) these terms vanish at $\bar{X}^i_{, \theta } =0$.
We thus conclude,
\begin{equation}
<< \bar{X}^4_{, \lambda \theta } \epsilon^\mu a A_\mu >> = <<O( \alpha )>>
\label{w21}
\end{equation}
Using
(\ref{w16a}),(\ref{w18}),(\ref{w16b}),(\ref{w21}),(\ref{w19}),(\ref{w16d}),(\ref{w17})
and (\ref{w16d}) in the successive terms on the right hand side of (\ref{w15})
we obtain,
\begin{equation}
\delta W[\bar{X}]|_{\bar{X}^i_{, \theta } =0=\bar{X}^i} = <<O(\frac{
1}{\eta} )>>
\label{w22}
\end{equation}

We could have alternatively considered the symmetry associated with
$S_{5\mu}$ transformations $ ( \delta \neq 0, \epsilon =0 )$ in (\ref{w9}).
In view of the overall $Sp(2)$ symmetry of the formulation, we will obtain
the analogous relation
\begin{equation}  
\delta W[\bar{X}]|_{X^i_{, \lambda }=0=X^i} = O( \frac{ 1}{\eta})
\label{w23}
\end{equation} 
The relations (\ref{w22}) and (\ref{w23}) are statements of
formal $ OSp(3,1| 2) $ symmetry as $\eta\rightarrow \infty $. These
contain in them the consequences of BRS and anti-BRS invariance. These
consequences can be obtained in a manner analogous to the argument following
Eq. (19) of the Ref. \cite{sp} ( See also \cite{bpm} for alternative
procedure for the entire derivation). They result in equations
\begin{eqnarray} 
\frac{ \partial W}{\partial \theta }|_{X^i_{, \theta }=0=X^i} &=& O( \frac{
1}{\eta}) 
\label{w24a} \\
\frac{ \partial W}{\partial \lambda }|_{X^i_{, \lambda }=0=X_i} &=& O(
\frac{ 1}{\eta})
\label{w24b}
\end{eqnarray} 
for BRS and anti-BRS symmetry respectively.

These equations can also be alternatively
verified evaluating $W[X]$ along the lines of Ref \cite{sdj}
and evaluating  $\frac{ \partial W}{\partial \theta }$ and $ \frac{
\partial W}{\partial \lambda }$ along the line of Ref \cite{zpc,sdj2} using BRS
/anti-BRS symmetry of the resultant $W$

\section{ PHYSICAL MEANING OF $OSp(3,1|2)$ INVARIANCE}

We expand the multiplet $\bar{A}_i (\bar{x}) $ explicitly as
\begin{eqnarray} 
\bar{A}_i (\bar{x}) &=& \left (\begin{array}{l}
A_\mu (\bar{x}) \\ c_4 (\bar{x}) \\ c_5 (\bar{x}) \end{array} \right )
\equiv \left ( \begin{array}{l}
A_\mu^T (\bar{x}) +\sqrt{ \frac{ \eta}{3}}A_\mu^L (\bar{x}) +\sqrt{ \frac{
\eta}{3}}A_\mu^t (\bar{x}) \\ \sqrt{ \frac{ \eta}{3}}\bar{c} (\bar{x}) \\
\sqrt{ \frac{ \eta}{3}}c (\bar{x}) \end{array} \right ) \nonumber \\  
&=& \sqrt{ \frac{ \eta}{3}} \left ( \begin{array}{l} A_\mu^L (\bar{x})
+A_\mu^t (\bar{x}) \\ \bar{c} (\bar{x}) \\ c (\bar{x}) \end{array} \right )
+ \left ( \begin{array}{l} A_\mu^T (\bar{x}) \\0\\ \end{array} \right )
\nonumber \\  
&\equiv& \sqrt{\frac{\eta}{3}} \left [\bar{A}_\mu^R (\bar{x}) +\sqrt{ \frac{
3}{\eta}} A_\mu^T (\bar{x}) \right ]
\end{eqnarray}
Here $\bar{A}_\mu^R (\bar{x}) $, in particular, contains the fields
corresponding to the set $R$

We further expand the transformation laws for fields under
$ OSp(3,1|2) $ viz.
\begin{eqnarray} 
A_\mu ^\prime (\bar{x}) &=& A_\mu (\bar{x}) -\delta _\mu b c_4 (\bar{x}) +
\epsilon _\mu a c_5 (\bar{x}) -( \epsilon a \lambda +\delta b \theta )^\nu
\partial _\nu A_\mu (\bar{x}) \nonumber \\  &-& \delta \cdot x b A_{\mu,
\lambda } (\bar{x}) + \epsilon \cdot x a A_{\mu, \theta } (\bar{x}) 
\nonumber \\  
c^ \prime _4 (\bar{x}) &=& c_4 (\bar{x}) + \epsilon ^\mu a A_\mu (\bar{x}) 
 -( \epsilon a \lambda + \delta b \theta )^\nu \partial _\nu c_4 \nonumber
\\  &-& \delta \cdot x b c_{4, \lambda } (\bar{x}) + \epsilon \cdot x c_{4,
\theta } (\bar{x}) \nonumber \\  
c_5^ \prime (\bar{x}) &=& c_5 (\bar{x}) + \delta ^\mu b A_\mu (\bar{x}) -(
\epsilon a \lambda +\delta b \theta )^\nu \partial _\nu c_5 \nonumber \\ 
&-& \delta \cdot x b c_{5, \lambda }(\bar{x}) + \epsilon \cdot x a c_{5,
\theta } (\bar{x}) 
\end{eqnarray}
in powers of $\eta $. We find that these read
 \begin{eqnarray} 
A_\mu^{R \prime } (\bar{x}) &=& A_\mu^R (\bar{x}) +P^R_{\mu\nu}[- \delta _\mu b \bar{c}
(\bar{x}) +\epsilon _\mu a c (\bar{x}) -( \epsilon a \lambda + \delta b
\theta )^\nu \partial _\nu A^R_\mu (\bar{x}) \nonumber \\  
&-& \delta \cdot x b A_{\mu, \lambda }^R (\bar{x}) + \epsilon \cdot x a
A^R_{\mu, \theta} (\bar{x})] + 0( \frac{ 1}{\sqrt{\eta}}) \nonumber \\ 
\bar{c}^\prime (\bar{x}) &=& \bar{c} (\bar{x}) + \epsilon a A_\mu ^R
(\bar{x}) -( \epsilon a \lambda + \delta b \theta )^\nu \partial _\nu
\bar{c} (\bar{x}) \nonumber \\  &-& \delta \cdot x b \bar{c}^R_{, \lambda
}(\bar{x}) + \epsilon \cdot x a \bar{c}_{, \theta } (\bar{x}) + 0( \frac{
1}{\sqrt{\eta}}) \nonumber \\ 
c^\prime (\bar{x}) &=& c (\bar{x}) + \delta ^\mu b A_\mu^R (\bar{x}) -(
\epsilon a \lambda + \delta b \theta )^\nu \partial _\nu c (\bar{x})
\nonumber \\  &-& \delta \cdot x b c_{, \lambda } (\bar{x}) + \epsilon
\cdot x a c_{, \theta } (\bar{x}) + 0( \frac{ 1}{\sqrt{\eta}})
\label{53}
\end{eqnarray} 
and
\begin{equation} 
A^{\prime T}_\mu (\bar{x}) = 
A_\mu ^T (\bar{x}) + 0(\sqrt{\eta})
\label{53a}
\end{equation}
[$P^R_{\mu\nu}$ is the projection operator that projects away the
transverse part]. 
We note that as $\eta\rightarrow\infty$, (\ref{53}) refers to the
transformations within the set $A^R$ only.

Thus, in the limit $\eta \rightarrow \infty $, the $ OSp(3,1|2) $ 
transformations, in particular, contain a set of symmetry transformations
among the members of the redundant set R. The WT identities are a
particular consequence of these symmetries. A special consequence
of the WT identities is the cancellation of the contributions from the set
R in the intermediate states in the unitarity relations using the Cutkowsky
rules \cite{tht} .

In the present superspace formulation, we have an explicit construction of
a set of symmetry transformations amongst this set R; originating from the
original $ OSp(3,1|2) $ transformation which, as we have shown, lead to WT
identities in particular. Thus, this formulation can be looked upon as an
explicit realization of that relationship that is expected to exist with in
the fields of the set R that is ultimately known to lead to mutual
cancellations in the cutting equations.

{\bf Acknowledgment:}

SDJ acknowledges the hospitality provided by the theory group,
Saha Institute of Nuclear Physics, Calcutta, India, where part
of the work was done.

\end{document}